\documentclass{mem}
\usepackage{natbib}\usepackage{txfonts}\usepackage{balance}
\usepackage{graphicx}
\usepackage[a4paper,breaklinks,dvipdfm]{hyperref}
\idline{75}{282}
\begin{document}
\def\teff{$T\rm_{eff }$}
\def\kms{$\mathrm {km s}^{-1}$}
\setlength{\textfloatsep}{2mm}
\setlength{\parskip}{1mm}

\title{Fundamental properties of T dwarfs from model fits to FIRE spectra}

\author{C. P. Nicholls\inst{1} 
\and A. J. Burgasser\inst{1}
\and C. V. Morley\inst{2}}

\institute{University of California San Diego, 9500 Gilman Dve, La Jolla, CA 92093, USA
\email{cnicholls@physics.ucsd.edu}
\and University of California Santa Cruz, 1156 High St, Santa Cruz, CA 95060, USA}

\authorrunning{Nicholls, Burgasser \& Morley}

\titlerunning{Fundamental properties of T dwarfs}

\abstract{We fit a sample of 49 R=6000 NIR (0.9 - 2.5 $\mu$m) T dwarf spectra obtained with Magellan's FIRE spectrograph with two different atmospheric model sets to compare the derived physical parameters such as \teff, $\log g$, cloud opacity, and rotational velocity between the models, as well as their reliability. Many of our T dwarfs have distance measurements, which allows us to calculate their radii during the fitting, which can be compared to evolutionary models to determine age, mass and potentially the presence of unseen companions.
We present our spectral sample and model fits, and comment on the measured fundamental properties of these T dwarfs. Our analysis allows us to identify global deviations between models and observed spectra, and hence provides important feedback for the next generation of substellar atmospheric models.
}
\maketitle{}

\section{Introduction}
\vspace{-2mm}
Determining the physical parameters for individual T dwarfs is paramount for characterising them as a population, testing evolutionary models, and using these long-lived objects to measure the chemical evolution and star formation history of the Galaxy. Unfortunately, the low luminosities of T dwarfs inhibits the acquisition of high resolution spectra, and persistent discrepancies between observations and models result in significant parameter uncertainties.  

We observed a sample of 49 L9-T9 dwarfs with the Magellan FIRE spectrograph \citep{simcoe08}, the largest yet sample of T dwarfs with medium-resolution (R=6000) NIR (0.9 - 2.5 $\mu$m) spectra. Most sources were selected to be within 20 pc of the Sun and have declinations below 15$^{\circ}$, but the sample also includes resolved binaries and spectral binary candidates \citep{burgasser10}. We derive fundamental parameters for our sample by fitting their spectra with two sets of atmospheric models: the BT-Settl models of \cite{bt-settl} and the sulfide cloud models of \cite{morley}. Using two model sets will allow us to compare the derived physical parameters and assess their reliability. Both grids span the expected temperature and gravity ranges of T dwarfs (\teff \ = 500 - 1300 K, $\log g$ = 4.0 - 5.5), and the Morley grid also spans several different levels of sulfide cloud opacity, represented by the cloud sedimentation parameter $f_{sed}$~=~2~- nc. This work is ongoing and the results presented here are preliminary. 

\vspace{-3mm}
\section{Model fitting method}
\vspace{-2mm}
All models were convolved with a Gaussian of FWHM = 50 \kms \ to represent FIRE's line spread function. Then each model was rotationally broadened to $v \sin i$ from 0 to 100 \kms \ in steps of 5 \kms. 

Each model was fit to each dwarf by shifting it to the observed radial velocity, desampling to the wavelengths of the spectrum, and scaling the model flux to match the observed spectrum by minimising the reduced $\chi^2$ between the spectrum and the model. All pixels received equal weighting except regions of strong telluric absorption between photometric bands (see Fig.~\ref{firstpass}), and bad pixels, which received zero weighting. The best fit to a given dwarf is the model with the lowest reduced $\chi^2$. 

\vspace{-5mm}
\section{First pass fit results}
\begin{figure*}[p]
\begin{center}
\includegraphics[width = 0.75\textwidth]{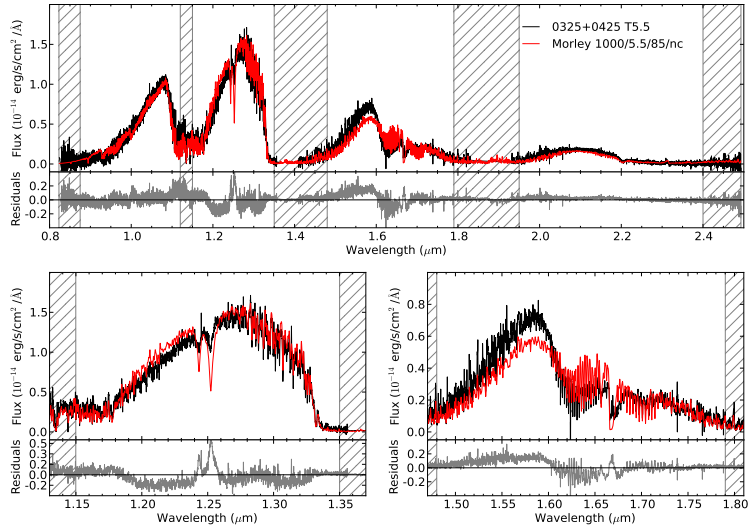}
\caption{Example best first-pass fit for T5.5 dwarf 0325+0425, which lies on the Morley model grid, with $\chi^2_{red} = 30.35$. In each plot, upper panel shows observed spectrum and best fit model, lower panel shows O-C residuals. Upper plot is the full NIR spectrum. Lower left plot is a closeup of the J band, lower right plot is a closeup of the H band. Grey hatching denotes wavelengths not included in the fit. Model parameters are shown in upper plot legend as \teff/$\log g$/$v \sin i$/$\rm{f_{sed}}$.}
\label{firstpass}
\end{center}
\end{figure*}

An example best fit is shown in Fig.~\ref{firstpass}. Note the large residuals from the two strong potassium lines at $\sim 1.24 \mu m$ and the methane features on the red side of the H band, $\sim 1.62 - 1.8 \mu m$. These two regions represent global deviations of the models, visible in model fits to most of our sample. We believe that the broad potassium lines drive the fits to unrealistically high $v \sin i$, as many dwarfs were fitted at the maximum $v \sin i$. We also know these spectral features have uncertain or incomplete opacities and/or treatment of broadening in the atmospheric models \citep{morley}. 

\vspace{-5mm}
\section{Second pass fit results}
\begin{figure*}[p]
\begin{center}
\includegraphics[width = 0.75\textwidth]{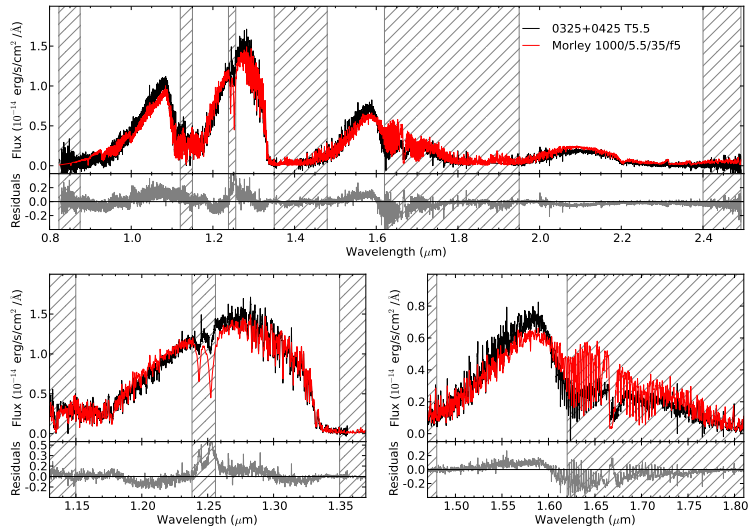}
\caption{Example best second-pass fit for T5.5 dwarf 0325+0425, $\chi^2_{red} = 12.40$. The absolute best second pass fit still lies on the Morley model grid but some parameters have changed.}
\label{secondpass}
\end{center}
\end{figure*}

A second pass fit was made by excluding the poor wavelength regions identified above. An example fit is shown in Fig.~\ref{secondpass}. Note that deviations appear lower for this second pass fit when the wavelengths used in both the first and second pass fits are compared. Second pass fit deviations are lower over the whole sample.

\begin{figure}
\begin{center}
\includegraphics[width = 0.5\textwidth]{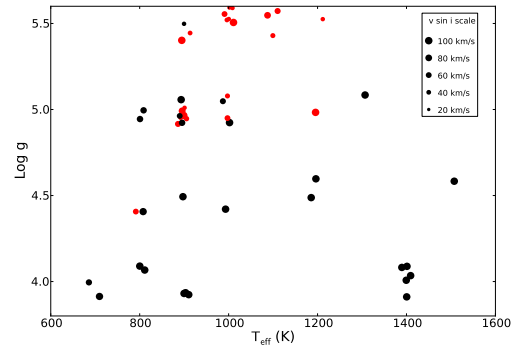}
\caption{$\log g$ vs \teff \ for second pass fits of the entire sample. Point size is relative to $v \sin i$. Black represents dwarfs best fit with a model with no sulfide clouds, red represents dwarfs best fit with a model containing some sulfide cloud opacity. Random scatter has been added so points can be differentiated.}
\label{loggteff}
\end{center}
\end{figure}

Fig.~\ref{loggteff} shows the parameters derived from the second pass fits for the entire sample. Nearly half the sample are best fit by a model with sulfide cloud opacity. This supports prior work indicating that including cloud physics in T dwarf atmospheric models is important to accurately recreate the spectra \citep{burgasseretal10,morley}.

\vspace{-6mm}
\section{Future work}
\vspace{-1mm}
The second pass fits are still far from perfect, and while this is partly limited by the models themselves, improving the fitting process further may have a significant effect on fit quality, and hence reliability of derived parameters. We are currently investigating whether any wavelength regions are sensitive to particular parameters, and hope to use the results of this to either reliably fix one or more parameters prior to fitting, or to influence a more sophisticated weighting scheme during fitting.

\vspace{-5mm}
\bibliographystyle{aa}
\bibliography{bdbiblio}

\begin{thebibliography}{5}
\expandafter\ifx\csname natexlab\endcsname\relax\def\natexlab#1{#1}\fi

\bibitem[{{Allard} \& {Freytag}(2010)}]{bt-settl}
{Allard}, F. \& {Freytag}, B. 2010, Highlights of Astronomy, 15, 756

\bibitem[{{Burgasser} {et~al.}(2010{\natexlab{a}}){Burgasser}, {Cruz},
  {Cushing}, {Gelino}, {Looper}, {Faherty}, {Kirkpatrick}, \&
  {Reid}}]{burgasser10}
{Burgasser}, A.~J., {Cruz}, K.~L., {Cushing}, M., {et~al.} 2010{\natexlab{a}},
  \apj, 710, 1142

\bibitem[{{Burgasser} {et~al.}(2010{\natexlab{b}}){Burgasser}, {Simcoe},
  {Bochanski}, {Saumon}, {Mamajek}, {Cushing}, {Marley}, {McMurtry}, {Pipher},
  \& {Forrest}}]{burgasseretal10}
{Burgasser}, A.~J., {Simcoe}, R.~A., {Bochanski}, J.~J., {et~al.}
  2010{\natexlab{b}}, \apj, 725, 1405

\bibitem[{{Morley} {et~al.}(2012){Morley}, {Fortney}, {Marley}, {Visscher},
  {Saumon}, \& {Leggett}}]{morley}
{Morley}, C.~V., {Fortney}, J.~J., {Marley}, M.~S., {et~al.} 2012, \apj, 756,
  172

\bibitem[{{Simcoe} {et~al.}(2008){Simcoe}, {Burgasser}, {Bernstein}, {Bigelow},
  {Fishner}, {Forrest}, {McMurtry}, {Pipher}, {Schechter}, \&
  {Smith}}]{simcoe08}
{Simcoe}, R.~A., {Burgasser}, A.~J., {Bernstein}, R.~A., {et~al.} 2008, in
  Society of Photo-Optical Instrumentation Engineers (SPIE) Conference Series,
  Vol. 7014, Society of Photo-Optical Instrumentation Engineers (SPIE)
  Conference Series

\end{thebibliography}

\end{document}